\documentclass[12pt,preprint]{aastex}

\shorttitle{VHE Observations of GRB Locations}
\shortauthors{Horan D., et al.}

\begin{document}
\bibliographystyle{plainnat}
\title{Very High Energy Observations of Gamma-Ray Burst Locations with the Whipple Telescope}

\author{D. Horan\altaffilmark{1,7},
R. W. Atkins\altaffilmark{2},
H. M. Badran\altaffilmark{3},
G. Blaylock\altaffilmark{4},
S. M. Bradbury\altaffilmark{5},
J. H. Buckley\altaffilmark{6},
K. L. Byrum\altaffilmark{7},
O. Celik\altaffilmark{8},
Y. C. K. Chow\altaffilmark{8},
P. Cogan\altaffilmark{9},
W. Cui\altaffilmark{10},
M. K. Daniel\altaffilmark{9},
I. de la Calle Perez\altaffilmark{11},
C. Dowdall\altaffilmark{9},
A. D. Falcone\altaffilmark{12},
D. J. Fegan\altaffilmark{9},
S. J. Fegan\altaffilmark{8},
J. P. Finley\altaffilmark{10},
P. Fortin\altaffilmark{13},
L. F. Fortson\altaffilmark{14},
G. H. Gillanders\altaffilmark{15},
J. Grube\altaffilmark{5},
K. J. Gutierrez\altaffilmark{6},
J. Hall\altaffilmark{2},
D. Hanna\altaffilmark{16},
J. Holder\altaffilmark{5},
S. B. Hughes\altaffilmark{6},
T. B. Humensky\altaffilmark{17},
G. E. Kenny\altaffilmark{15},
M. Kertzman\altaffilmark{18},
D. B. Kieda\altaffilmark{2},
J. Kildea\altaffilmark{16},
H. Krawczynski\altaffilmark{6},
F. Krennrich\altaffilmark{19},
M. J. Lang\altaffilmark{15},
S. LeBohec\altaffilmark{2},
G. Maier\altaffilmark{5},
P. Moriarty\altaffilmark{20},
T. Nagai\altaffilmark{19},
R. A. Ong\altaffilmark{8},
J. S. Perkins\altaffilmark{6},
D. Petry\altaffilmark{21},
J. Quinn\altaffilmark{9},
M. Quinn\altaffilmark{20},
K. Ragan\altaffilmark{16},
P. T. Reynolds\altaffilmark{22},
H. J. Rose\altaffilmark{5},
M. Schroedter\altaffilmark{19},
G. H. Sembroski\altaffilmark{10},
D. Steele\altaffilmark{14},
S. P. Swordy\altaffilmark{17},
J. A. Toner\altaffilmark{15},
L. Valcarcel\altaffilmark{16},
V. V. Vassiliev\altaffilmark{8},
R. G. Wagner\altaffilmark{7},
S. P. Wakely\altaffilmark{17},
T. C. Weekes\altaffilmark{1},
R. J. White\altaffilmark{5},
D. A. Williams\altaffilmark{23}}

\email{deirdreh@hep.anl.gov}

\altaffiltext{1}{Fred Lawrence Whipple Observatory, Harvard-Smithsonian Center for Astrophysics, P.O. Box 97, Amado, AZ 85645-0097}
\altaffiltext{2}{Physics Department, University of Utah, Salt Lake City, UT 84112, USA}
\altaffiltext{3}{Department of Physics, Tanta University, Tanta, Egypt}
\altaffiltext{4}{Department of Physics, University of Massachusetts, Amherst, MA 01003-4525, USA}
\altaffiltext{5}{School of Physics and Astronomy, University of Leeds, Leeds, LS2 9JT, UK}
\altaffiltext{6}{Department of Physics, Washington University, St. Louis, MO 63130, USA}
\altaffiltext{7}{Argonne National Laboratory, 9700 S. Cass Avenue, Argonne, IL 60439, USA}
\altaffiltext{8}{Department of Physics and Astronomy, University of California, Los Angeles, CA 90095, USA}
\altaffiltext{9}{School of Physics, University College Dublin, Belfield, Dublin 4, Ireland}
\altaffiltext{10}{Department of Physics, Purdue University, West Lafayette, IN 47907, USA}
\altaffiltext{11}{Department of Physics, University of Oxford, Oxford, OX1 3RH, UK}
\altaffiltext{12}{Department of Astronomy and Astrophysics, 525 Davey Lab., Penn. State University, University Park, PA 16802, USA}
\altaffiltext{13}{Department of Physics and Astronomy, Barnard College, Columbia University, NY 10027, USA}
\altaffiltext{14}{Astronomy Department, Adler Planetarium and Astronomy Museum, Chicago, IL 60605, USA}
\altaffiltext{15}{Physics Department, National University of Ireland, Galway, Ireland}
\altaffiltext{16}{Physics Department, McGill University, Montreal, QC H3A 2T8, Canada}
\altaffiltext{17}{Enrico Fermi Institute, University of Chicago, Chicago, IL 60637, USA}
\altaffiltext{18}{Department of Physics and Astronomy, DePauw University, Greencastle, IN 46135-0037, USA}
\altaffiltext{19}{Department of Physics and Astronomy, Iowa State University, Ames, IA 50011, USA}
\altaffiltext{20}{Department of Physical and Life Sciences, Galway-Mayo Institute of Technology, Dublin Road, Galway, Ireland}
\altaffiltext{21}{N.A.S.A./Goddard Space-Flight Center, Code 661, Greenbelt, MD 20771, USA}
\altaffiltext{22}{Department of Applied Physics and Instrumentation, Cork Institute of Technology, Bishopstown, Cork, Ireland}
\altaffiltext{23}{Santa Cruz Institute for Particle Physics and Department of Physics, University of California, Santa Cruz, CA 95064, USA}

\begin{abstract} 

Gamma-ray burst (GRB) observations at very high energies (VHE,
{\it{E}} $>$ 100\,GeV) can impose tight constraints on some GRB
emission models. Many GRB afterglow models predict a VHE component
similar to that seen in blazars and plerions, in which the GRB
spectral energy distribution has a double-peaked shape extending into
the VHE regime. VHE emission coincident with delayed X-ray flare
emission has also been predicted. GRB follow-up observations have had
high priority in the observing program at the Whipple 10\,m Gamma-ray
Telescope and GRBs will continue to be high priority targets as the
next generation observatory, VERITAS, comes on-line. Upper limits on
the VHE emission, at late times ($>$\,$\sim$\,4 hours), from seven
GRBs observed with the Whipple Telescope are reported here.

\end{abstract}


\section{Introduction}
\label{intro}

Since their discovery in 1969 \citep{Klebesadel:73}, gamma-ray bursts
(GRBs) have been well studied at many wavelengths. Although various
open questions remain on their nature, there is almost universal
agreement that the basic mechanism is an expanding relativistic
fireball, that the radiation is beamed, that the prompt emission is
due to internal shocks and that the afterglow arises from external
shocks. It is likely that Lorentz factors of a few hundred are
involved, with the radiating particles, either electrons or protons,
being accelerated to very high energies.  GRBs are sub-classified into
two categories, long and short bursts, based on the timescale over
which 90\% of the prompt gamma-ray emission is detected.

Recently, the Swift GRB Explorer \citep{Gehrels:04} has revealed that
many GRBs have associated X-ray flares \citep{Burrows:05,
Falcone:06a}. These flares have been detected between 10$^2$\,s and
10$^5$\,s after the initial prompt emission and have been found to
have fluences ranging from a small fraction of, up to a value
comparable to, that contained in the prompt GRB emission. This X-ray
flare emission has been postulated to arise from a number of different
scenarios, including late central engine activity where the GRB
progenitor remains active for some time after, or re-activates after,
the initial explosion
\citep{Kumar:00,Zhang:06,Nousek:06,Perna:05,Proga:06,King:05} and
refreshed shocks which occur when slower moving shells ejected by the
central engine in the prompt phase catch up with the afterglow shock
at late times \citep{Rees:98,Sari:00,Granot:03a,Guetta:06}. For short
GRBs, shock heating of a binary stellar companion has also been
proposed \citep{MacFadyen:05}. It is not yet clear whether the X-ray
flares are the result of prolonged central engine activity, refreshed
shocks or some other mechanism \citep{Panaitescu:06}. A very high
energy (VHE; E $>$\,100\,GeV) component of this X-ray flare emission
has also been predicted \citep{Wang:06}.

Within the standard fireball shock scenario
\citep{Rees:92,Meszaros:93,Sari:98}, many models have been proposed
which predict emission at and above GeV energies during both the
prompt and afterglow phases of the GRB. These have been summarized by
\citet{Zhang:04}, and references therein, and include leptonic models
in which gamma rays are produced by electron self-inverse-Compton
emission from the internal shocks or from the external forward or
reverse shocks. Other models predict gamma rays from proton
synchrotron emission or photomeson cascade emission in the external
shock or from a combination of proton synchrotron emission and
photomeson cascade emission from internal shocks.

Although GRB observations are an important component of the program at
many VHE observatories, correlated observations at these short
wavelengths remain sparse even though tantalizing and inherently very
important. The sparsity of observations of GRBs at energies above
10\,MeV is dictated not by lack of interest in such phenomena, or the
absence of theoretical predictions that the emission should occur, but
by the experimental difficulties.

For the observation of photons of energies above 100\,GeV, only
ground-based telescopes are available at present. These ground-based
telescopes fall into two broad categories, air shower arrays and
atmospheric Cherenkov telescopes (of which the majority are Imaging
Atmospheric Cherenkov Telescopes, or IACTs). The air shower arrays,
which have wide fields of view making them particularly suitable for
GRB searches, are relatively insensitive. There are several reports
from these instruments of possible TeV emission.
\citet{Padilla:98:Airobicc} reported possible VHE emission at {\it{E}}
$>$ 16 TeV from GRB\,920925c. While finding no individual burst which
is statistically significant, the Tibet-AS$\gamma$ Collaboration found
an indication of 10\,TeV emission in a stacked analysis of 57 bursts
\citep{Amenomori:01:TibetGRBs}. The Milagro Collaboration reported on
the detection of an excess gamma-ray signal during the prompt phase of
GRB\,970417a with the Milagrito detector \citep{Atkins00}. In all of
these cases however, the statistical significance of the detection is
not high enough to be conclusive. In addition to searching the Milagro
data for VHE counterparts for 25 satellite-triggered GRBs
\citep{Atkins:05:MilagroGRBCounterparts}, the Milagro Collaboration
conducted a search for VHE transients of 40 seconds to 3 hours
duration in the northern sky \citep{Atkins:04:MilagroGRB}. No evidence
for VHE emission was found from either of these searches and upper
limits on the VHE emission from GRBs were derived. Atmospheric
Cherenkov telescopes, particularly those that utilize the imaging
technique, are inherently more flux-sensitive than air shower arrays
and have better energy resolution but are limited by their small
fields of view (3\,-\,5$^\circ$) and low duty cycle ($\sim$\,7\%). In
the Burst And Transient Source Explorer \citep{Meegan:92} era
(1991\,-\,2000), attempts at GRB monitoring were limited by slew times
and uncertainty in the GRB source position \citep{Connaughton:97}. 

Swift, the first of the next generation of gamma-ray satellites which
will include AGILE (Astro-rivelatore Gamma a Immagini LEggero) and the
Gamma-ray Large Area Space Telescope (GLAST), is beginning to provide
arcminute localizations so that IACTs are no longer required to scan a
large GRB error box in order to achieve full coverage of the possible
emission region. The work in this paper covers the time period prior
to the launch of the Swift satellite.

The minimum detectable fluence with an IACT, such as the Whipple 10m,
in a ten second integration is $<$\,10$^{-8}$\,erg\,cm$^{-2}$ (5
photons of 300\,GeV in 5\,x\,10$^8$\,cm$^2$ collection area). This is
a factor of $>$\,100 better than GLAST will achieve (3 photons of
10\,GeV in 10$^4$\,cm$^2$ collection area). This ignores the large
solid angle advantage of a space telescope and the possible steepening
of the observable spectrum because of the inherent emission mechanism
and the effect of intergalactic absorption by pair production. There
have been many predictions of high energy GRB emission in and above
the GeV energy range
\citep{Meszaros:94a,Boettcher:98,PillaLoeb:98,Wang:01,Zhang:01,Guetta:03b,Dermer:04,Fragile:04};
also see \citet{Zhang:04} and references therein.

Until AGILE and GLAST are launched, the GRB observations that were
made by the EGRET experiment on the Compton Gamma Ray Observatory
(CGRO) will remain the most constraining in the energy range from
30\,MeV to 30\,GeV. Although EGRET was limited by a small collection
area and large dead time for GRB detection, it made sufficient
detections to indicate that there is a prompt component with a hard
spectrum that extends at least to 100\,MeV energies. The average
spectrum of four bursts detected by EGRET (GRBs\,910503, 930131,
940217 \& 940301) did not show any evidence for a cutoff up to 10 GeV
\citep{Dingus:01}. The relative insensitivity of EGRET was such that
it was not possible to eliminate the possibility that all GRBs had
hard components \citep{Dingus98}. EGRET also detected an afterglow
component from GRB\,940217 that extended to 18\,GeV for at least 1.5
hours after the prompt emission indicating that a high-energy spectral
component can extend into the GeV band for a long period of time, at
least for some GRBs \citep{Hurley94}. The spectral slope of this
component is sufficiently flat that its detection at still higher
energies may be possible \citep{Mannheim96}. \citet{Meszaros:94b}
attribute this emission to the combination of prompt MeV radiation
from internal shocks with a more prolonged GeV inverse Compton
component from external shocks. It is also postulated that this
emission could be the result of inverse Compton scattering of X-ray
flare photons \citep{Wang:06}. Although somewhat extreme parameters
must be assumed, synchrotron self-Compton emission from the reverse
shock is cited as the best candidate for this GeV emission by
\citet{Granot:03b}, given the spectral slope that was recorded. This
requirement of such extreme parameters naturally explains the lack of
GRBs for which such a high energy component has been
observed. \citet{Guetta:03a} postulate that some GRB explosions occur
inside pulsar wind bubbles. In such scenarios, afterglow electrons
upscatter pulsar wind bubble photons to higher energies during the
early afterglow thus producing GeV emission such as that observed in
GRB\,940217.

The GRB observational data are extraordinarily complex and there is no
complete and definitive explanation for the diversity of properties
observed. It is important to establish whether there is, in general, a
VHE component of emission present during either the prompt or
afterglow phase of the GRB. Understanding the nature of such emission
will provide important information about the physical conditions of
the emission region.  One definitive observation of the prompt or
afterglow emission could significantly influence our understanding of
the processes at work in GRB emission and its aftermath.

In this paper, the GRBs observed with the Whipple 10\,m Gamma-ray
Telescope in response to HETE-2 and INTEGRAL notifications are
described. The search for VHE emission is restricted to times on the
order of hours after the GRB. In Section~\ref{observe}, the observing
strategy, telescope configuration, and data analysis methods used in
this paper are described. The properties of the GRBs observed and
their observation with the Whipple Telescope are described in
Section~\ref{grbs}. Finally, in Section~\ref{resultsAndDiscussion},
the results are summarized and their implications discussed in the
context of some theoretical models that predict VHE emission from
GRBs. The sensitivity of future instruments such as VERITAS to GRBs is
also discussed.

\section{The Gamma-ray Burst Observations}
\label{observe}

\subsection{Telescope Configuration}

The observations presented here were made with the 10\,m Gamma-ray
Telescope at the Fred Lawrence Whipple Observatory.  Constructed in
1968, the telescope has been operated as an IACT since 1982
\citep{Kildea:06}. In September 2005, the observing program at the
10\,m was redefined and the instrument was dedicated solely to the
monitoring of TeV blazars and the search for VHE emission from
GRBs. Located on Mount Hopkins approximately 40 km south of Tucson in
southern Arizona at an altitude of 2300\,m, the telescope consists of
248 hexagonal mirror facets mounted on a 10\,m spherical dish with an
imaging camera at its focus. The front-aluminized mirrors are mounted
using the Davies-Cotton design \citep{Davies57}.

The imaging camera consists of 379 photo-multiplier tubes (PMTs)
arranged in a hexagonal pattern. A plate of light-collecting cones is
mounted in front of the PMTs to increase their light-collection
efficiency. A pattern-sensitive trigger \citep{Bradbury:02}, generates
a trigger whenever three adjacent PMTs register a signal above a level
preset in the constant fraction discriminators. The PMT signals for
each triggering event are read out and digitized using
charge-integrating analog to digital converters. In this way, a map of
the amount of charge in each PMT across the camera is recorded for
each event and stored for offline analysis. The telescope triggers at
a rate of $\sim\,25$\,Hz (including background cosmic ray triggers)
when pointing at high ($>$\,50$^\circ$) elevation. Although sensitive
in the energy range from 200\,GeV to 10\,TeV, the peak response energy
of the telescope to a Crab-like spectrum during the observations
reported upon here was approximately 400\,GeV. This is the energy at
which the telescope is most efficient at detecting gamma rays and is
subject to a 20\% uncertainty.

\subsection{Observing Strategy}

Burst notifications at the Whipple Telescope for the observations
described here were received via email from the Global Coordinates
Network \citep{GCN:webpage}. When a notification email arrived, the
GRB location and time were extracted and sent to the telescope
tracking control computer. An audible alarm sounded to alert the
observer of the arrival of a burst notification. If at sufficient
elevation, the observer approved the observations and the telescope
was commanded to slew immediately to the location of the GRB. The
Whipple Telescope slews at a speed of 1$^\circ$\,s$^{-1}$ and
therefore can reach any part of the visible sky within three minutes.

Seven different GRB locations were observed with the Whipple 10\,m
Telescope between November 2002 and April 2004. These observations are
summarized in Table~\ref{grb_summary}. At the time these data were
taken, the point spread function of the Whipple Telescope was
approximately 0.1$^\circ$ which corresponds to the field of view of
one PMT. The positional offsets for the GRB observations (see
Table~\ref{observations_and_results}) were all less than this so a
conventional ``point source'' analysis was performed.

\subsection{Data Analysis}

The data were analyzed using the imaging technique and analysis
procedures pioneered and developed by the Whipple Collaboration
\citep{Reynolds93}. In this method, each image is first cleaned to
exclude the signals from any pixels that are most likely the result of
noise. The cleaned images are then characterized by calculating and
storing the first, second, and third moments of the light distribution
in each image. The parameters and this procedure are described
elsewhere \citep{Reynolds93}. Since gamma-ray images are known to be
compact and elliptical in shape, while those generated by cosmic ray
showers tend to be broader with more fluctuations, cuts can be derived
on the above parameters which reject approximately 99.7\% of the
background images while retaining over 50\% of those generated by
gamma-ray showers. These cuts are optimized using data taken on the
Crab Nebula which is used as the standard candle in the TeV sky.

Two different modes of observation are employed at the Whipple
Telescope, ``{\it{On\,-\,Off}}'' and ``{\it{Tracking}}''
\citep{Catanese:98}. The choice of mode depends upon the nature of the
target. The GRB data presented here were all taken in the
{\it{Tracking}} mode. Unlike data taken in the {\it{On\,-\,Off}} mode,
scans taken in the {\it{Tracking}} mode do not have independent
control data which can be used to establish the background level of
gamma-ray like events during the scan. These control data are
essential in order to estimate the number of events passing all cuts
which would have been detected during the scan in the absence of the
candidate gamma-ray source. In order to perform this estimate, a
tracking ratio is calculated by analyzing ``darkfield data''
{\citep{Horan02}}. These consist of {\it{Off-source}} data taken in
the {\it{On\,-\,Off}} mode and of observations of objects found not to
be sources of gamma rays. A large database of these scans is analyzed
and in this way, the background level of events passing all gamma-ray
selection criteria can be characterized as a function of zenith
angle. Since the GRB data described in this paper were taken at
elevations between 50$^\circ$ and 80$^\circ$, a large sample of
darkfield data ($\sim$ 233 hours) spanning a similar zenith angle
range was analyzed so that the background during the gamma-ray burst
data runs could be estimated.

\section{The Gamma-ray Bursts}
\label{grbs}

This paper concentrates on the GRB observations made in response to
HETE-2 and INTEGRAL triggers with the Whipple 10\,m Gamma-ray
Telescope; observations made in response to Swift triggers are the
subject of a separate paper \citep{Dowdall}. When the GRB data were
filtered to remove observations made at large zenith angles, during
inferior weather conditions, and of positions later reported to be the
result of false triggers or to have large positional errors, the data
from observations of seven GRB locations remained. These GRBs took
place between UT dates 021112 and 040422; two have redshifts derived
from spectral measurements, one has an estimated redshift and four lie
at unknown distances. Five of the sets of GRB follow-up observations
were carried out in response to triggers from the high energy
transient explorer 2 (HETE-2; \citet{Lamb:00}) while two sets of
observations were triggered by the international gamma-ray
astrophysical laboratory (INTEGRAL; \citet{Winkler:99}). In the
remainder of this section, the properties of each of the GRBs observed
and the results of these observations are presented. A summary of the
GRB properties is given in Table~\ref{grb_summary} while the
observations taken at the Whipple Observatory are summarized in
Table~\ref{observations_and_results}.

\subsection{GRB\,021112}

This was a long GRB with a duration of $>$ 5\,s and a peak flux of
$>$\,3\,x\,10$^{-8}$\,erg\,cm$^{-2}$\,s$^{-1}$ in the 8\,-\,40\,keV
band \citep{GCN1682:hete2}. In the 30\,-\,400\,keV energy band, the
burst had a peak energy of 57.15\,keV, a duration of 6.39s and a
fluence of 2.1\,x\,10$^{-7}$\,erg\,cm$^{-2}$ \citep{hete2webpage}. The
triggering instrument was the French Gamma Telescope (FREGATE)
instrument on HETE-2. The Milagro data taken during the time of this
burst were searched for GeV/TeV gamma-ray emission. No evidence for
prompt emission was found and a preliminary analysis, assuming a
differential photon spectral index of -2.4, gave an upper limit on the
fluence at the 99.9\% confidence level of {\it{J}}(0.2\,-\,20\,TeV)
$<$\,2.6\,x\,10$^{-6}$\,erg\,cm$^{-2}$ over a 5 second interval
\citep{McEnery:02:GCN1724}. Optical observations with the 0.6-meter
Red Buttes Observatory Telescope beginning 1.8 hours after the burst
did not show any evidence for an optical counterpart and placed a
limiting magnitude of {\it{R$_c$}}=21.8 (3\,sigma) on the optical
emission: at the time, this was the deepest non-detection of an
optical afterglow within 2.6 hours of a GRB
\citep{GCN1776:Optical021112}.

Two sets of observations on the location of GRB\,021112 were made with
the Whipple 10\,m Telescope. The first observations commenced 4.2
hours after the GRB occurred and lasted for 110.6
minutes. Observations were also taken for 55.3 minutes on the
following night, 28.6 hours after the GRB occurred. Upper limits
(99.7\% c.l.) of 0.20 Crab\footnote{Since the Crab is the standard
candle in the VHE regime, it is customary to quote upper limits as a
fraction of the Crab flux at the same energy.}  and 0.30 Crab (E $>$
400 GeV), respectively, were derived for these observations assuming a
Crab-like spectrum (spectral index of -2.49).

\subsection{GRB\,021204}

Little information is available in the literature on this HETE-2
burst. The GRB location was observed with a number of optical
telescopes (the RIKEN 0.2m \citep{Torii:02:GCN1730}, the 32 inch
Tenagra II \citep{Nysewander:02:GCN1735}, and the 1.05\,m Schmidt at
Kiso Observatory \citep{GCN1747:RIKEN}) but no optical transient was
found to a limiting magnitude of {\it{R}}=16.5, 2.1 hours after the
burst \citep{Torii:02:GCN1730}, and to {\it{R}}=18.8, 6.2 hours after
the burst \citep{GCN1747:RIKEN}.

Whipple observations of this burst location commenced 16.9 hours after
the GRB occurred and lasted for 55.3 minutes. An upper limit (99.7\%
c.l.) of 0.33 Crab was derived for the VHE emission above 400\,GeV
during these observations.

\subsection{GRB\,021211}

This long, bright burst was detected by all three instruments on
HETE-2. It had a duration $>$ 5.7\,s in the 8\,-\,40\,keV band with a
fluence of $\sim$ 10$^{-6}$\,erg\,cm$^{-2}$ during that interval
\citep{GCN1734:hete2}. The peak flux was
$>$\,8\,x\,10$^{-7}$\,erg\,cm$^{-2}$\,s$^{-1}$ (i.e. $>$\,20 Crab
flux) in 5\,ms \citep{GCN1734:hete2}. This burst had a peak energy of
45.56\,keV, a duration of 2.80s, and a fluence of
2.4\,x\,10$^{-6}$\,erg\,cm$^{-2}$ in the 30\,-\,400\,keV energy band
\citep{hete2webpage}.  \citet{Fox:EarlyOptical:03} reported on the
early optical, near-infrared, and radio observations of this
burst. They identified a break in the optical light curve of the burst
at t=0.1\,-\,0.2\,hr, which was interpreted as the signature of a
reverse shock. The light curve comprised two distinct phases. The
initial steeply-declining flash was followed by emission declining as
a typical afterglow with a power-law index close to 1.  KAIT
observations of the afterglow also detected the steeply declining
light curve and evidence for an early break
\citep{Li:EarlyOptical:03}. The optical transient was detected at many
observatories \citep{GCN1736:Super-LOTIS, GCN1737:KAIT,
GCN1744:ARC-ACO, GCN1750:MMT, GCN1751:Magellan-Baade}. The optical
transient faded from an R-band magnitude of 18.3, 20.7 minutes after
the burst, to an R-band magnitude of 21.1, 5.7 hours after the burst
\citep{Fox:EarlyOptical:03}. \citet{GCN1785:Redshift021211} derived a
redshift of 1.006 for this burst based on spectroscopic observations
carried out with the European Southern Observatory's Very Large
Telescope (VLT) at Paranal, Chile. Milagro searched for emission at
GeV/TeV energies over the burst duration reported by the HETE-2 wide
field X-ray monitor. They did not find any evidence for prompt
emission and a preliminary analysis, assuming a differential photon
spectral index of -2.4, gave an upper limit on the fluence at the
99.9\% confidence level of
{\it{J}}(0.2\,-\,20\,TeV)$<$3.8\,x\,10$^{-6}$\,erg\,cm$^{-2}$ over a 6
second interval \citep{GCN1740:Milagro}.

Whipple observations on this GRB location were initiated 20.7 hours
after the GRB and lasted for 82.8 minutes. An upper limit (99.7\%
c.l.) on the VHE emission of 0.33 Crab (E $>$ 400\,GeV) was derived
from these observations.

\subsection{GRB\,030329}

This GRB is one of the brightest bursts on record. It triggered the
FREGATE instrument on HETE-2 in the 6\,-\,120\,keV energy band. It had
a duration 22.76 seconds, a fluence of
1.1\,x\,10$^{-4}$\,erg\,cm$^{-2}$ and a peak energy of 67.86\,keV in
the 30\,-\,400\,keV band \citep{hete2webpage}. The peak flux over 1.2
seconds was 7\,x\,10$^{-6}$\,erg\,cm$^{-2}$\,s$^{-1}$ which is
$>$\,100 times the Crab flux in that energy band
\citep{GCN1997:hete2}.

The optical transient was identified by \citet{GCN1985:SSO}. Due to
its slow decay \citep{GCN1989:Kyoto} and brightness
({\it{R}}\,$\sim$\,13), extensive photometric observations were
possible, making this one of the best-observed GRB afterglows to
date. Early observations with the VLT \citep{GCN2020:VLT} revealed
evidence for narrow emission lines from the host galaxy indicating
that this GRB occurred at a low redshift of
{\it{z}}\,=\,0.1687. Observations of the afterglow continued for many
nights as it remained bright with a slow but uneven rate of decline
and exhibited some episodes of increasing brightness. These
observations are well-documented in the GCN archives. Spectral
measurements made on 6 April 2003 by \citet{GCN2107:MMT} showed the
development of broad peaks in flux, characteristic of a
supernova. Over the next few nights, the afterglow emission faded and
the features of the supernova became more prominent
\citep{Stanek:03}. These observations provided the first direct
spectroscopic evidence that at least a subset of GRBs is associated
with supernovae.

The afterglow was detected at many other wavelengths. Radio
observations with the VLA detected a 3.5\,mJy source at
8.46\,GHz. This is the brightest radio afterglow detected to date
\citep{GCN2014:VLA}. The afterglow was also bright at submillimeter
\citep{GCN2088:submm} and near infrared wavelengths
\citep{GCN2040:nir}. The X-ray afterglow was detected by RXTE during a
27-minute observation that began 4 hours 51 minutes after the burst
\citep{GCN1996:rxte}. The flux was
$\sim$1.4\,x\,10$^{-10}$\,erg\,cm$^{-2}$\,s$^{-1}$ in the
2\,-\,10\,keV band ($\sim$\,0.007\% of the Crab).

Whipple observations of the location of GRB\,030329 commenced 64.6
hours after the prompt emission. In total, 241.4 minutes of
observation were taken spanning five nights. The upper limits (99.7\%
c.l.) from each night of observation are listed in
Table~\ref{observations_and_results} and are displayed on the same
temporal scale as the optical light curve of the GRB afterglow in
Figure~\ref{GRB030329-lightcurve}. When these data were combined, an
upper limit (99.7\% c.l.) for the VHE emission above 400\,GeV of 0.17
Crab was derived.

\subsection{GRB\,030501}

This burst was initially detected by the imager on board the INTEGRAL
satellite (IBIS/ISGRI) and was found to have a duration of $\sim$\,40
seconds \citep{GCN2183:integral}. The burst was also detected by the
Ulysses spacecraft and the spectrometer instrument (SPI-ACS) on
INTEGRAL \citep{GCN2187:ipn}. Triangulation between these two
detections allowed a position annulus to be computed for this GRB. As
observed by Ulysses, it had a duration of $\sim$\,75 seconds and had a
25\,-\,100\,keV fluence of approximately
1.1\,x\,10$^{-6}$\,erg\,cm$^{-2}$ with a peak flux of
4.9\,x\,10$^{-7}$\,erg\,cm$^{-2}$\,s$^{-1}$ over 0.25
seconds. Follow-up optical observations with several telescopes did
not find evidence for an optical transient \citep{GCN2201:Wise,
GCN2202:crao, GCN2224:tarot} to a limiting magnitude of {\it{R}}=18.0,
0.3\,-\,17 minutes after the burst \citep{GCN2224:tarot} and to a
limiting magnitude of {\it{R}}=20.0, 16.5 hours after the burst
\citep{GCN2201:Wise}.

Whipple observations of this burst location commenced 6.6 hours after
its occurrence and continued for 83.1 minutes. An upper limit (99.7\%
c.l.) on the VHE emission (E $>$ 400\,GeV) during these observations
of 0.27 Crab was derived.

\subsection{GRB\,031026}

This burst was located by the FREGATE instrument on HETE-2. It had a
duration of 114.2 seconds with a fluence of
2.3\,x\,10$^{-6}$\,erg\,cm$^{-2}$ in the 25\,-\,100\,keV energy band
\citep{GCN2429:hete2} while in the 30\,-\,400\,keV energy band it had
a duration of 31.97\,s and a fluence of
2.8\,x\,10$^{-6}$\,erg\,cm$^{-2}$ \citep{hete2webpage}. Follow-up
optical observations were carried out with a number of instruments
including the 1.05\,m Schmidt at the Kiso Observatory
\citep{GCN2427:kiso}, the 32 inch Tenagra II Telescope
\citep{GCN2428:tenagra}, and the 1.0m Telescope at the Lulin
Observatory \citep{GCN2436:lulin}, but no optical transient was found
to a limiting magnitude of {\it{R}}=20.9 for observations taken
6\,-\,12 hours after the burst \citep{GCN2436:lulin} and to
{\it{I$_c$}}=20.4 from observations taken 3.9 and 25.7 hours after the
burst \citep{GCN2433:tenagra}. The 30m IRAM Telescope was used to
search the field around the GRB location but did not detect any source
with a 250\,GHz flux density $>$ 16\,mJy \citep{GCN2440:iram}. A
spectral analysis of the prompt X-ray and gamma-ray emission from this
burst revealed it to have a very hard spectrum which is unusual for
such a long and relatively faint burst \citep{GCN2432:hete2}. It was
noted that the counts ratio of $>$1.8 between the 7\,-\,30\,keV and
7\,-\,80\,keV FREGATE energy bands was one of the most extreme
measured \citep{GCN2429:hete2}. A ``new pseudo-redshift'' of
6.67\,$\pm$\,2.9 was computed for this burst using the prescription of
\citet{Pelangeon:06}.

Whipple observations of this burst location were initiated 3.7 minutes
after receiving the GRB notification. The burst notification however,
was not received until more than 3 hours after the prompt GRB
emission. Although Whipple observations commenced 3.3 hours after the
prompt emission, the first data run is not included here due to
inferior weather conditions. The data presented here commenced 3.7
hours after the GRB and continued for 82.7 minutes. An upper limit
(99.7\% c.l.) for the VHE emission ({\it{E}}\,$>$\,400\,GeV) of 0.41
Crab was derived.

\subsection{GRB\,040422}

This burst was detected by the imaging instrument (IBIS/ISGRI) on the
INTEGRAL satellite in the 15\,-\,200 keV energy band. It had a
duration of 8 seconds, a peak flux between 20 and 200\,keV of
2.7\,photons\,cm$^{-2}$\,s$^{-1}$ and a fluence (1\,s integration
time) of 2.5\,x\,10$^{-7}$\,erg\,cm$^{-2}$
\citep{GCN2572:integral}. Follow-up observations were carried out by
many groups but no optical transient was detected \citep{GCN2573:eso,
GCN2574:miyazaki, GCN2576:rotse, GCN2577:lulin, GCN2578:loiana,
GCN2580:crao, GCN2581:xinglong}. The ROTSE-IIIb Telescope at McDonald
Observatory began taking unfiltered optical data 22.1\,s after the
GRB. Using the first 110\,s of data, a limiting magnitude of
$\sim$\,17.5 was placed on the R-band emission from the GRB at this
time \citep{GCN2576:rotse}.

Whipple observations of this burst commenced 4.0 hours after the
prompt emission and continued for 27.6 minutes. An upper limit (99.7\%
c.l.) on the VHE emission (E $>$ 400\,GeV) of 0.62 Crab was derived.

\section{Results and Discussion}
\label{resultsAndDiscussion}

\begin{figure}
\resizebox*{0.45\textwidth}{!}{\includegraphics[draft=false]{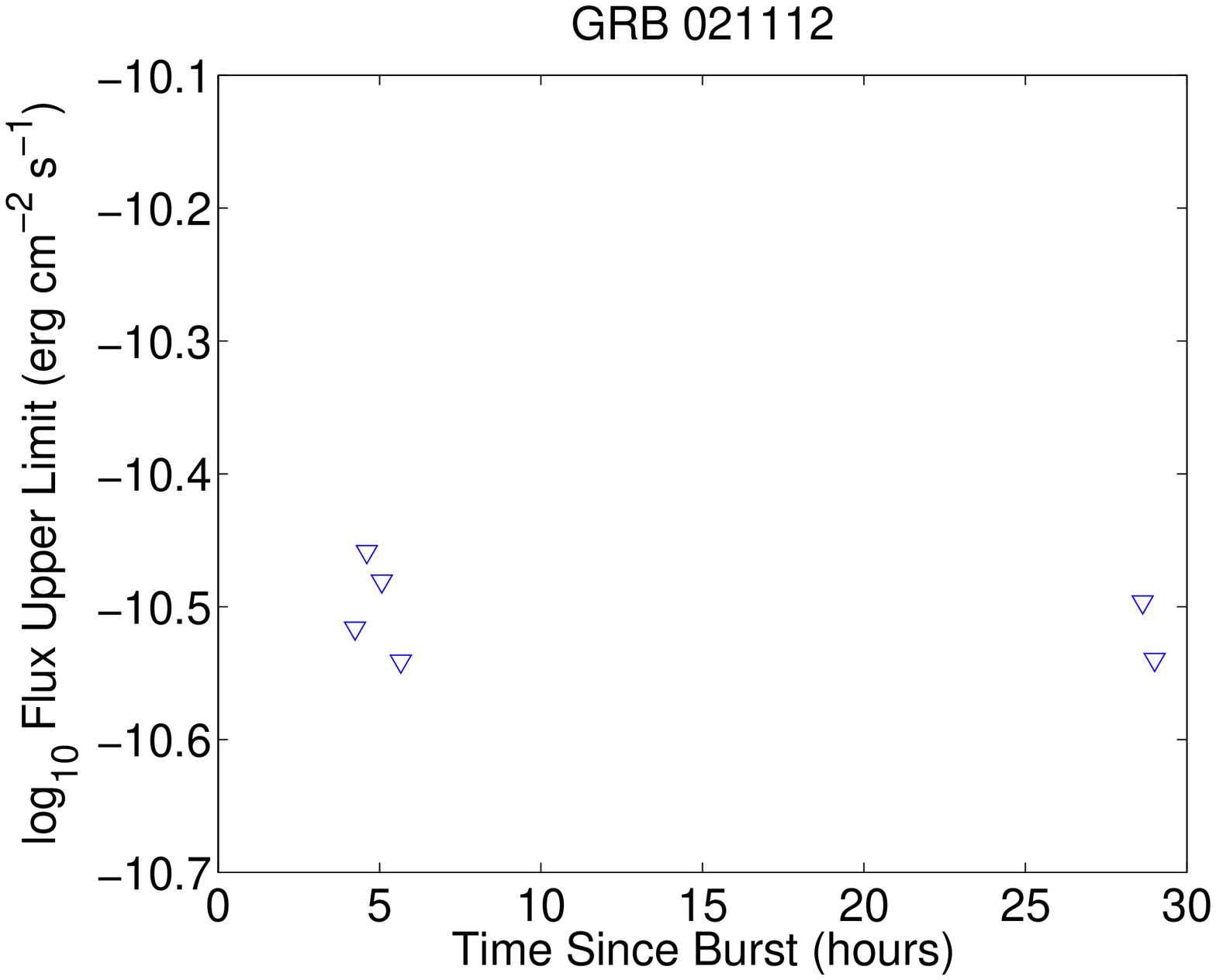}}%
\resizebox*{0.45\textwidth}{!}{\includegraphics[draft=false]{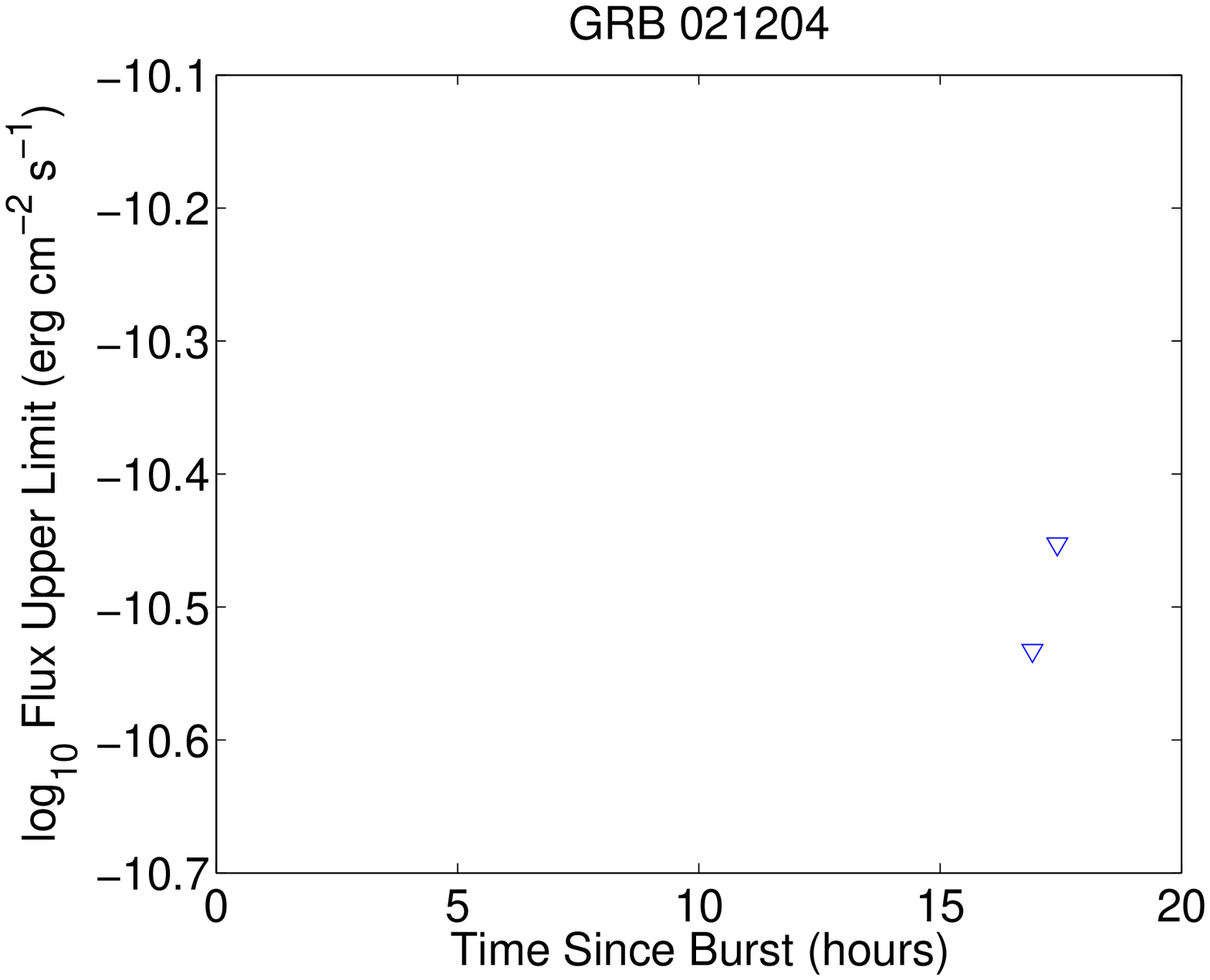}}\\[2ex]%
\resizebox*{0.45\textwidth}{!}{\includegraphics[draft=false]{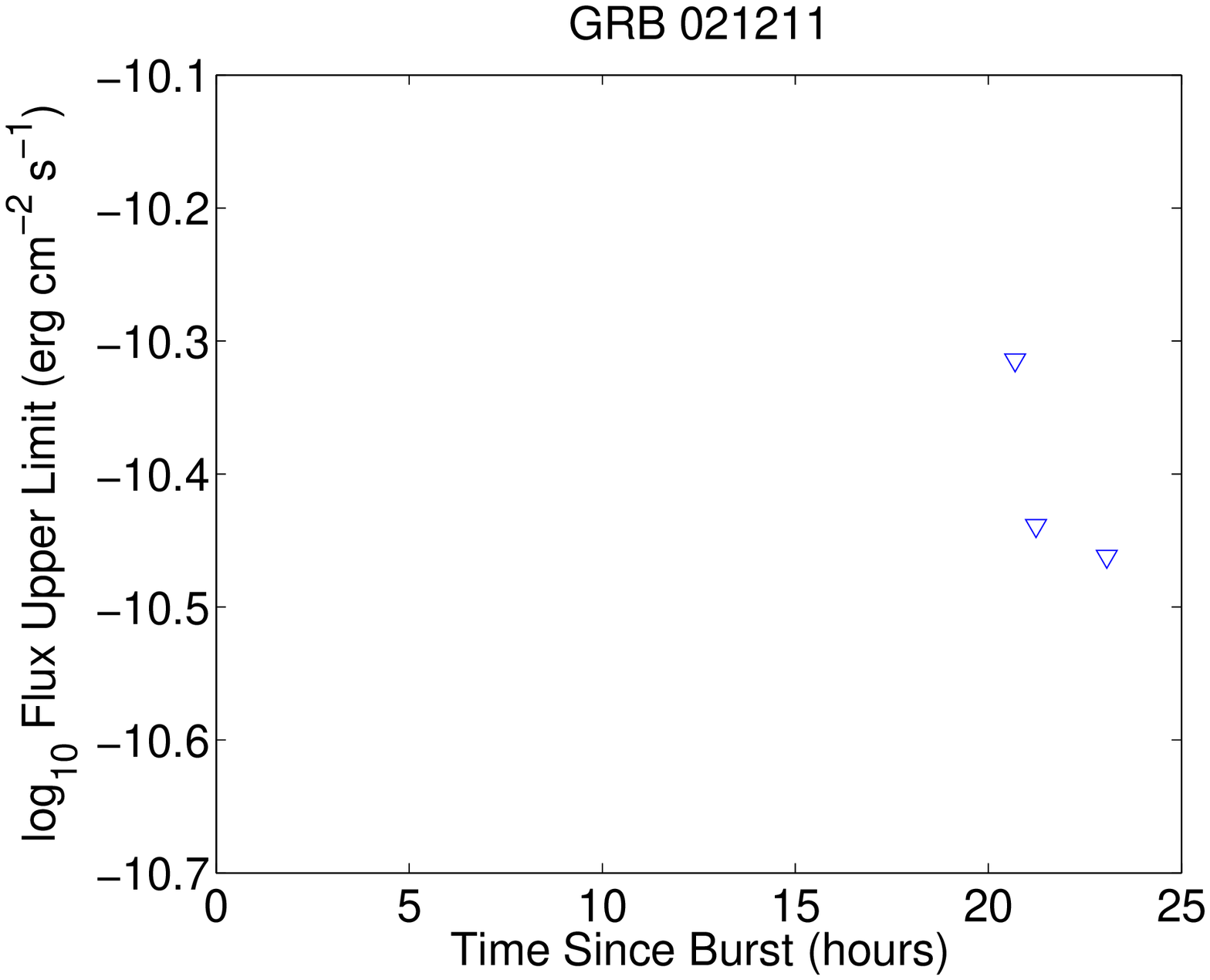}}%
\resizebox*{0.45\textwidth}{!}{\includegraphics[draft=false]{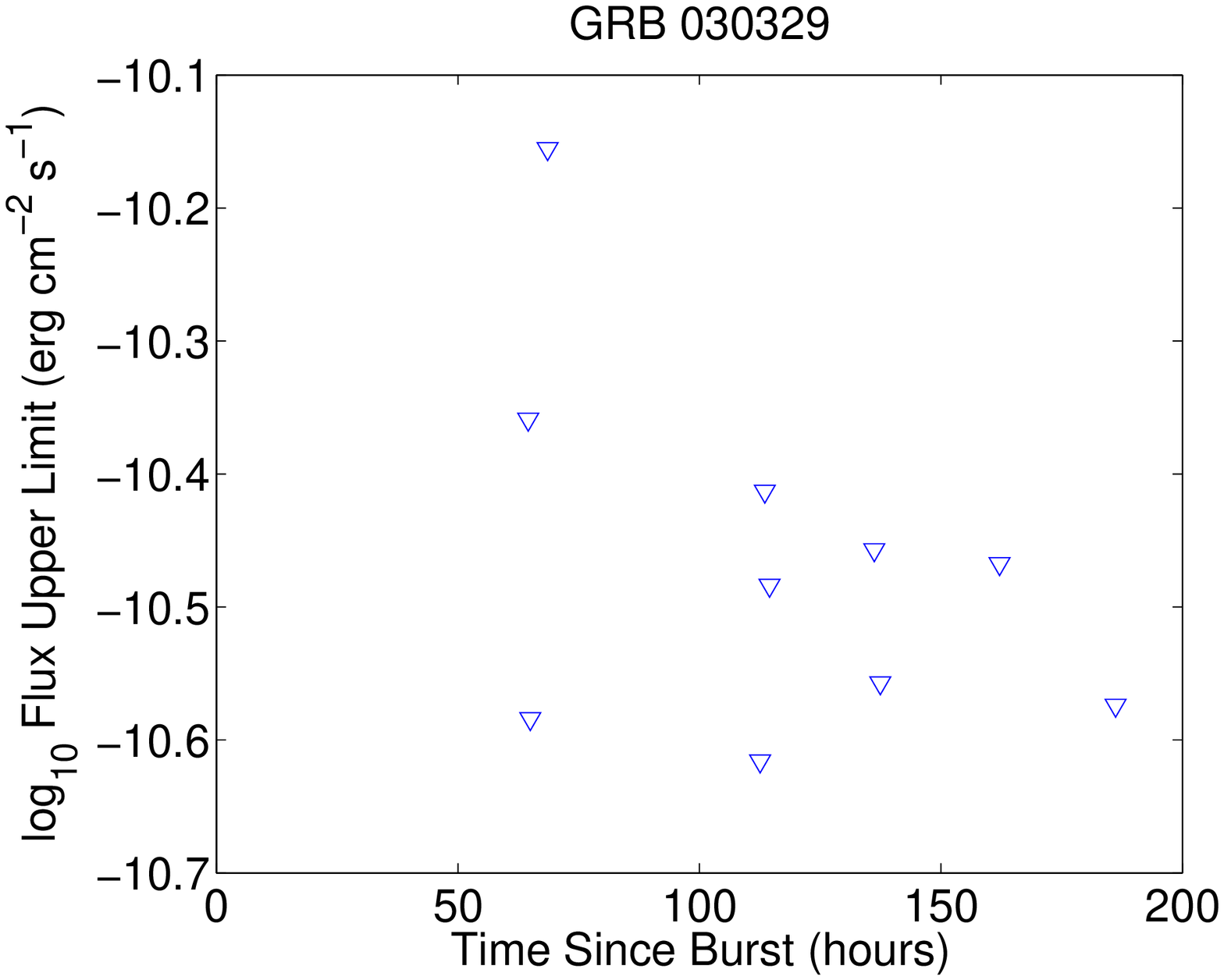}}\\[2ex]%
\resizebox*{0.45\textwidth}{!}{\includegraphics[draft=false]{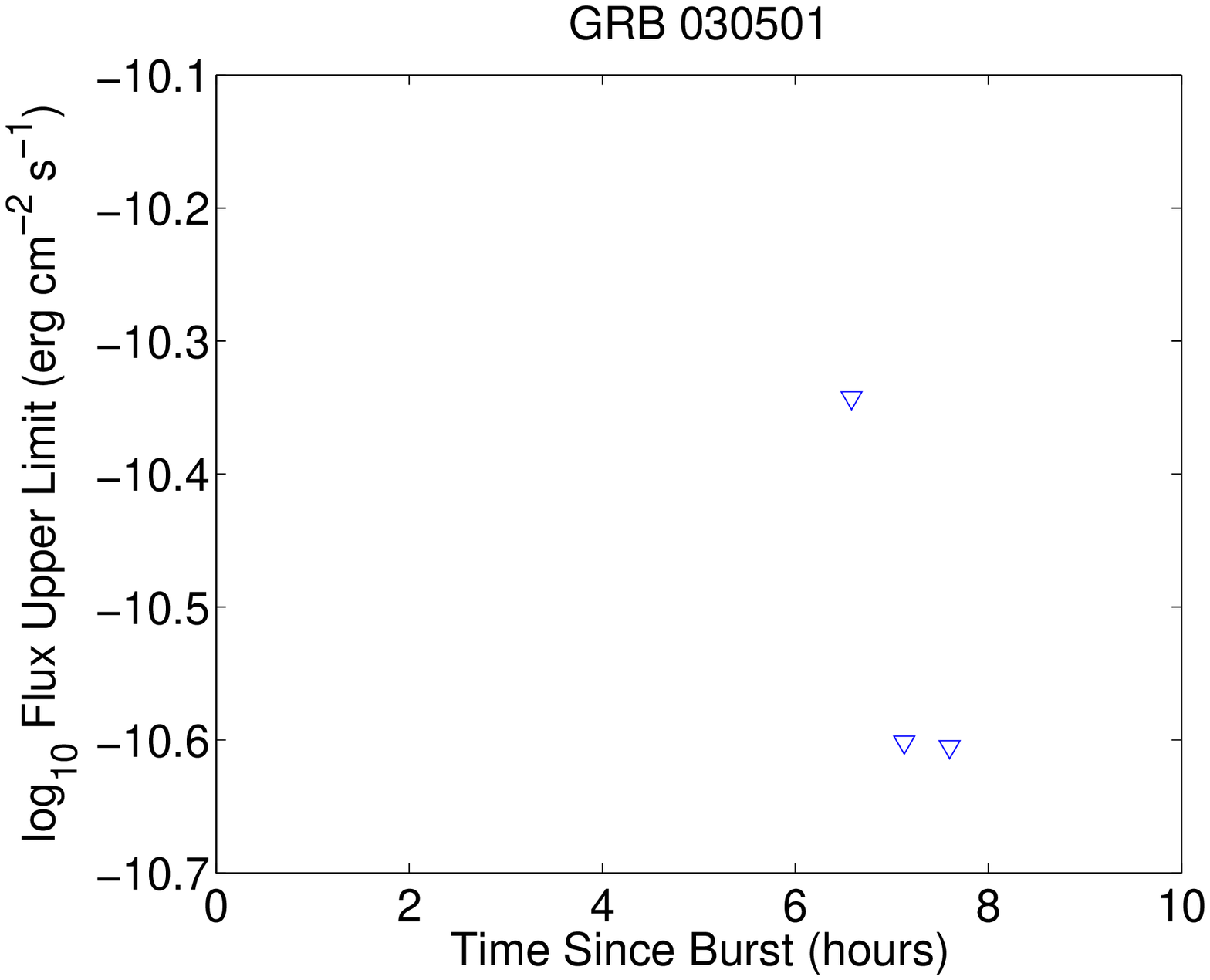}}%
\resizebox*{0.45\textwidth}{!}{\includegraphics[draft=false]{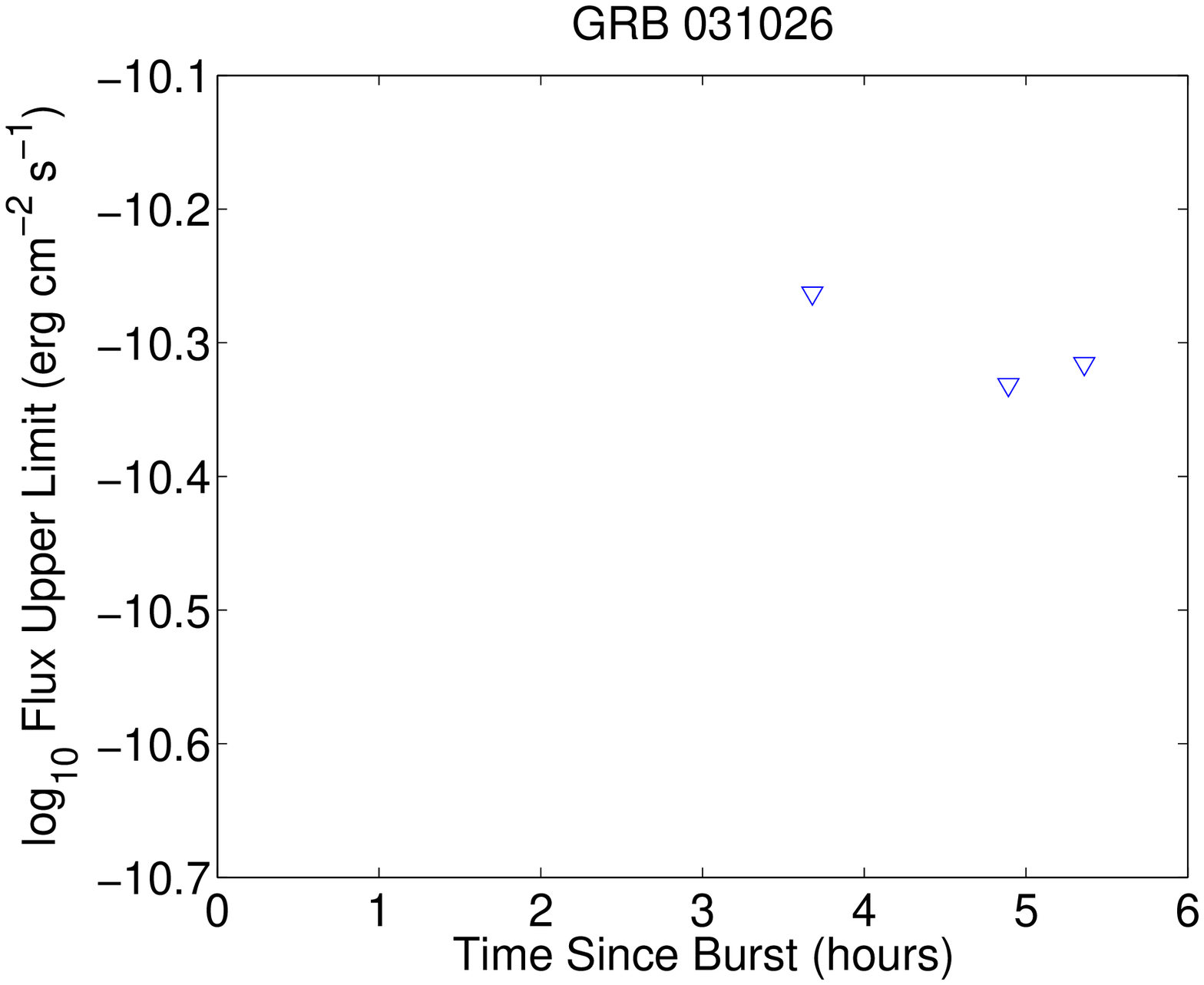}}
\caption{\label{FIG::UPPERLIMITS}For each GRB location observed, flux
upper limits in units of $10^{-11}$\,erg\,cm$^{-2}$\,s$^{-1}$ were
calculated for each 28-minute scan taken. These are plotted here as a
function of the time since the GRB prompt emission for each GRB. Only
one 28-minute observation was made on GRB\,040422 so the plot for this
GRB is not shown.}
\end{figure}

Upper limits on the VHE emission from the locations of seven GRBs have
been derived over different timescales. For each GRB, a number
(1\,-\,10) of follow-up 28-minute duration observations were taken
with the Whipple 10\,m Telescope. These GRB data were grouped by UT
day and were combined to give one upper limit for each day of
observation. The limits range from 20\% to 62\% of the Crab flux above
400\,GeV and are presented in Table~\ref{observations_and_results}. In
addition to calculating upper limits on the GRB emission for each day,
upper limits were calculated for each of the 28-minute scans. These
are plotted for each of the GRBs in Figure~\ref{FIG::UPPERLIMITS}.

The usefulness of the upper limits presented here is limited by the
fact that five of the GRBs occurred at unmeasured redshifts thus
making it impossible to infer the effects of the infrared background
light on those observations. In addition to this, the earliest
observation was not made at Whipple until 3.68 hours after the prompt
GRB emission. Although the Whipple 10\,m Telescope is capable of
beginning GRB observations less than 2 minutes after receiving
notification, a number of factors, including notifications arriving
during daylight and delays in the distribution of the GRB locations,
delayed the commencement of the GRB observations presented
here. Although data-taking for GRB\,031026 began 3.7 minutes after the
GRB notification was received, this notification was not distributed
by the GCN until 3.3 hours after the GRB had occurred. Thus, the
observations presented here cannot be used to place constraints on the
VHE component of the initial prompt GRB emission and pertain only to
the afterglow emission and delayed prompt emission from GRBs.

One of the main obstacles for VHE observations of GRBs is the distance
scale. Pair production interactions of gamma rays with the infrared
photons of the extragalactic background light attenuate the gamma-ray
signal thus limiting the distance over which VHE gamma rays can
propagate. Recently however, the H.E.S.S. telescopes have detected the
blazar PG\,1553+113 \citep{Aharonian06:PG1553}. The redshift of this
object is not known but there are strong indications that it lies at
{\it{z}}\,$>$\,0.25, possibly as far away as {\it{z}}\,=\,$0.74$. This
could represent a large increase in distance to the most distant
detected TeV source, revealing more of the universe to be visible to
TeV astronomers than was previously thought. Although GRBs lie at
cosmological distances, many have been detected at redshifts
accessible to VHE observers. Of the GRBs studied here, only 2 had
spectroscopic redshifts measured while the redshift of one was
estimated by \citet{Pelangeon:06} using an improved version of the
redshift estimator of \citet{Atteia03}. Since all of the GRBs
discussed here were long bursts, it is likely that their redshifts are
of order 1. Due to the unknown redshifts of most of the bursts and the
uncertainty in the density of the extragalactic background light, the
effects of the absorption of VHE gamma rays by the infrared background
light have not been included here.

\begin{figure}
\resizebox*{0.5\textwidth}{!}{\includegraphics[draft=false]{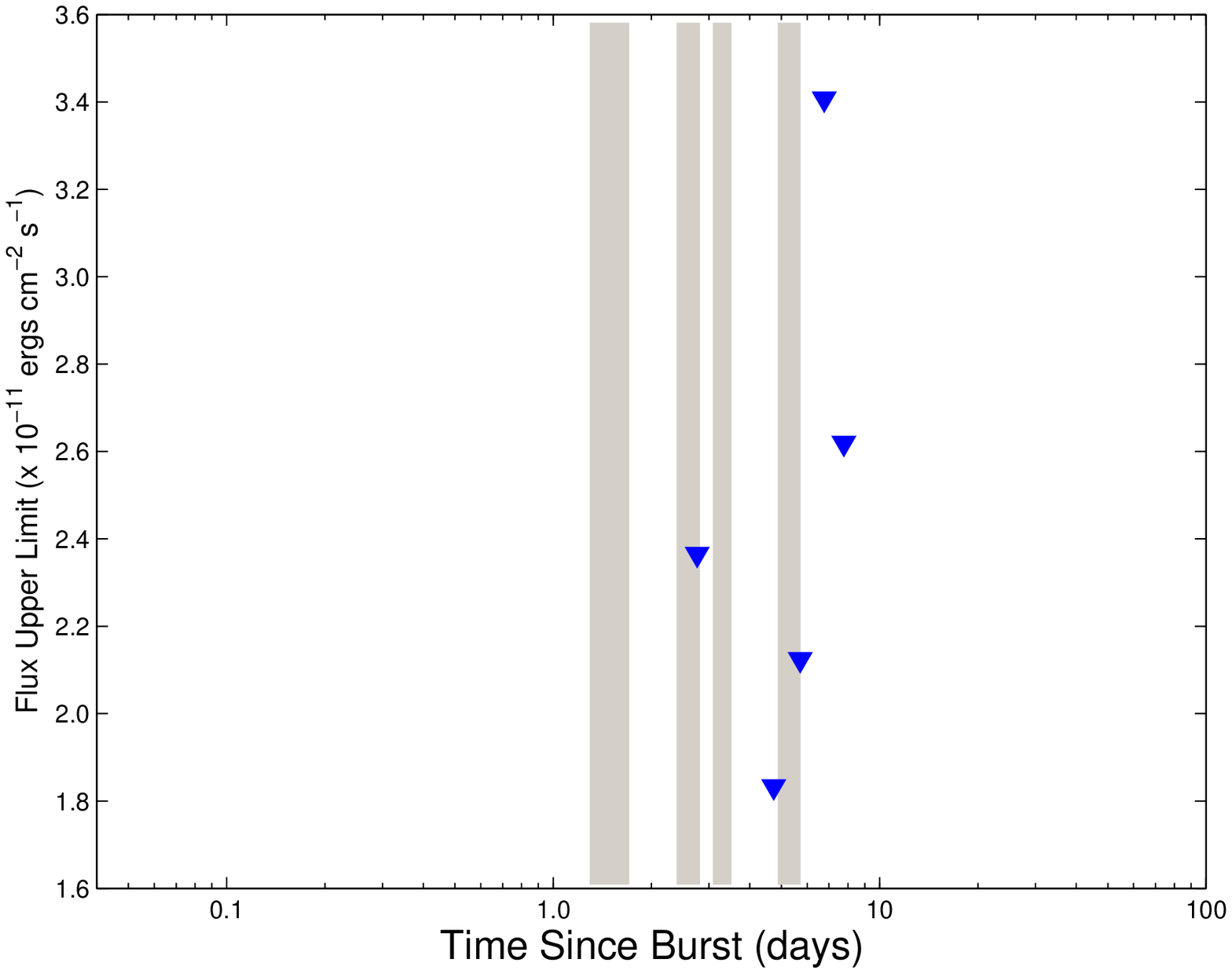}}\\%
\resizebox*{0.5\textwidth}{!}{\includegraphics[draft=false]{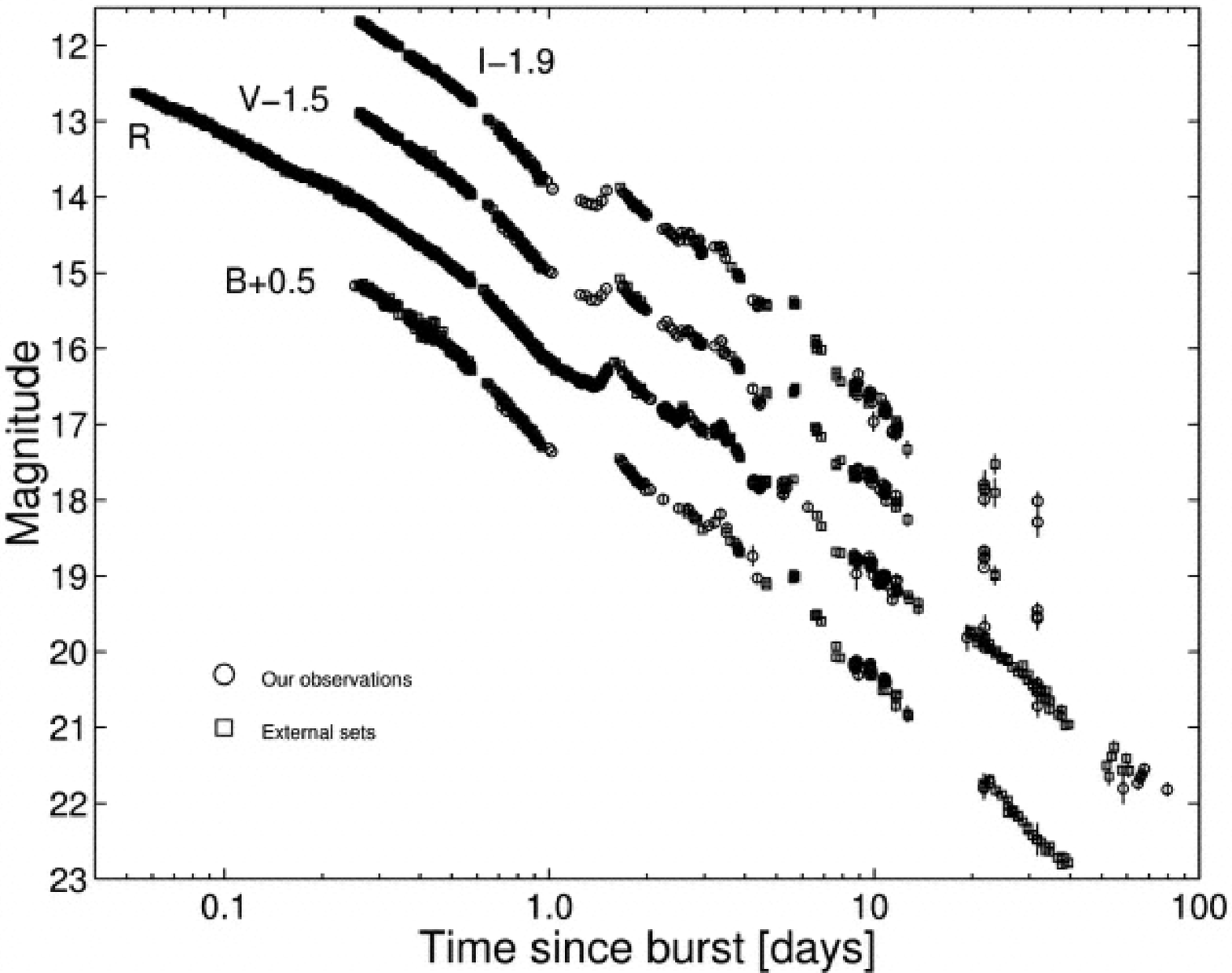}}%
\caption{\label{GRB030329-lightcurve}Top: The flux upper limits above
  400\,GeV (99.7\% c.l.) on the VHE emission from GRB\,030329. The
  time periods during which the four bumps in the lightcurve occur
  \citep{Granot:03a} are shown as shaded rectangles. Bottom: The
  optical light curve of GRB\,030329 taken from
  \citet{Lipkin:OpticalAG:04}. The time since the GRB is shown with
  the same scale on the x-axis of both plots.}
\end{figure}

\citet{Granot:03a} analyzed the late time light curve of GRB\,030329
and find that the large variability observed at several times
(t=1.3-\,$\sim$\,1.7 days, $\sim$\,2.4\,-\,2.8 days,
$\sim$\,3.1\,-\,3.5 days and at $\sim$\,4.9\,-\,5.7 days) after the
burst is most likely the result of refreshed shocks. These time
intervals have been highlighted in the top panel of
Figure~\ref{GRB030329-lightcurve} and it can be seen that some of the
observations taken at Whipple occurred during these times thus
imposing upper limits on the VHE emission during these refreshed
shocks. Since GRB\,030329 occurred at a low redshift ($z\,=\,$0.1685),
it is possible that the effects of infrared absorption on any VHE
emission component may not have been significant enough to absorb all
VHE photons over the energy range to which Whipple is sensitive.

\begin{figure}
\epsscale{1}
\resizebox*{0.5\textwidth}{!}{\includegraphics[draft=false]{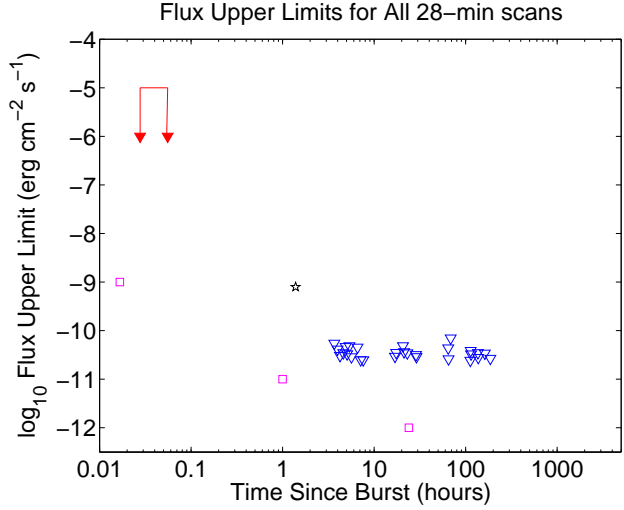}}%
\caption{\label{ulTimeLog}The flux upper limits above 400\,GeV for all
of the GRBs observed (blue triangles). The limits are plotted as a
function of time since the GRB prompt emission. The approximate flux
level at 400\,GeV predicted by \citet{Peer:SED:04} is indicated by the
red solid lines along with the time interval during which it is
predicted to occur; magenta squares show the emission at 400\,GeV
predicted by \citet{Zhang:01} at various times after the GRB prompt
emission; the prediction of \citet{Guetta:03a} for VHE emission at
250\,GeV 5 x 10$^3$\,s after the burst from the combination of
external Compton and synchrotron self-Compton emission is shown by the
black star.}
\end{figure}

Figure~\ref{ulTimeLog} shows these scan-by-scan upper limits as a
function of time since the prompt GRB emission. Also plotted are the
predicted fluxes at various times after the GRB by \citet{Zhang:01}
and \citet{Peer:SED:04} at $\sim$ 400\,GeV, and by \citet{Guetta:03a}
at 250\,GeV. Although the peak response energy of the Whipple
Telescope at the time of these observations was 400\,GeV, it still had
sensitivity, albeit somewhat reduced, at 250\,GeV.

\citet{Razzaque04} predict a delayed GeV component in the GRB
afterglow phase from the inverse-Compton up-scattering on external
shock electrons. The duration of such a component is predicted to be
up to a few hours, softening with time. \citet{Zhang:01} investigated
the different radiation mechanisms in GRB afterglows and identified
parameter-space regimes in which different spectral components
dominate. They found that the inverse-Compton GeV photon component is
likely to be significantly more important than a possible proton
synchrotron or electron synchrotron component at these high energies.
The predictions of \citet{Zhang:01} for VHE emission at different
times after a typical ``Regime II'' burst are shown by squares on
Figure~\ref{ulTimeLog}. Although the observations presented here do
not constrain these predictions, the sensitivity is close to that
required to detect the emission predicted.

A recent analysis of archival data from the EGRET calorimeter has
found a multi-MeV spectral component in the prompt phase of
GRB\,941017, that is distinct from the lower energy component
\citep{Gonzalez03}.  This high energy component appeared between 10\,s
and 20\,s after the start of the GRB and had a roughly constant flux
with a relatively hard spectral slope for $\sim$\,200\,s. This
observation is difficult to explain within the standard synchrotron
model, thus indicating the existence of new phenomena.
\citet{Granot:03b} investigated possible scenarios for this high
energy spectral component and found that most models fail. They
concluded that the best candidate for the emission mechanism is
synchrotron self-Compton emission from the reverse shock and predicted
that a bright optical transient, similar to that observed in
GRB\,990123, should accompany this high energy
component. \citet{Peer:SED:04} explain this high energy tail as
emission from the forward shock electrons in the early afterglow
phase. These electrons inverse-Compton scatter the optical photons
that are emitted by the reverse shock electrons resulting in powerful
VHE emission for 100\,s to 200\,s after the burst as indicated by the
lines on Figure~\ref{ulTimeLog}. Although the observations presented
here did not commence early enough after the prompt GRB emission to
constrain such models, the sensitivity of the Whipple Telescope is
such that the VHE emission predicted by these models would be easily
detectable for low redshift bursts.

The prediction of \citet{Guetta:03a} for VHE emission 5 x 10$^3$\,s
after the burst from the combination of external Compton emission (the
relativistic electrons behind the afterglow shock upscatter the
plerion radiation) and synchrotron self-Compton emission (the
electrons accelerated in the afterglow emit synchrotron emission and
then upscatter this emission to the VHE regime) is indicated by a star
on Figure~\ref{ulTimeLog}. The emission is predicted to have a cutoff
at $\sim$\,250\,GeV due to pair production of the high energy photons
with the radiation field of the pulsar wind bubble. For afterglows
with an external density similar to that of the inter-stellar medium,
photons of up to 1\,TeV are possible. It can be seen that, although
the upper limits presented here are below the predicted flux from
\citet{Guetta:03a}, the observations at Whipple took place after this
emission was predicted to have occurred. Had data taking at Whipple
commenced earlier, the emission predicted by these authors should have
been detectable for nearby GRBs.

\citet{Razzaque04} investigated the interactions of GeV and higher
energy photons in GRB fireballs and their surroundings for the prompt
phase of the GRB. They predict that high energy photons escaping from
the fireball will interact with infrared and microwave background
photons to produce delayed secondary photons in the GeV\,-\,TeV range.
Although observations of the prompt phase of GRBs are difficult with
IACTs since they are pointed instruments with small fields of view
which must therefore be slewed to respond to a burst notification,
observations in time to detect the delayed emission are possible.

There are many emission models which predict significant VHE emission
during the afterglow phase of a GRB either related to the afterglow
emission itself or as a VHE component of the X-ray flares that have
been observed in many Swift bursts. \citet{OBrien:06} analyzed 40
Swift bursts which had narrow-field instrument data within 10 minutes
of the trigger and found that $\sim$\,50\% had late (t\,$>$\,T90)
X-ray flares. If the bulk of the radiation comes via synchrotron
radiation as is usually supposed, then by analogy with other systems
with similar properties (supernova remnants, active galactic nuclei
jets), it is natural to suppose that there must also be an inverse
Compton component by which photons are boosted into the GeV\,-\,TeV
energy range. This process is described by \citet{PillaLoeb:98} who
discuss the relationship between the energy at which the high energy
cutoff occurs, the bulk Lorentz factor and the size of the emission
region. A high energy emission component due to inverse Compton
emission has also been considered in detail for GRB afterglows by
\citet{SariEsin:01}; the predicted flux at GeV\,-\,TeV energies is
comparable to that near the peak of the radiation in the afterglow
synchrotron spectrum. Only direct observations can confirm whether
this is so. \citet{Guetta:03b} predict that the $\sim$300 GeV photons
from the prompt GRB phase will interact with background IR photons,
making delayed high energy emission undetectable unless the
intergalactic magnetic fields are extremely small.

The Swift GRB Explorer has shown that $\sim$\,50\% of GRBs have one or
more X-ray flares. These flares have been detected up to 10$^5$\,s
($\sim$\,28 hours) after the prompt emission
\citep{Burrows:05}. Indeed, the delayed gamma-ray component detected
in BATSE bursts \citep{Connaughton:02} may also be associated with
this phenomenon.  Recently, \citet{Wang:06} have predicted VHE
emission coincident in time with the X-ray flare photons. In this
model, if the X-ray flares are caused by late central engine activity,
the VHE photons are produced from inverse Compton scattering of the
X-ray flare photons from forward shock electrons. If the X-ray flares
originate in the external shock, VHE photons can be produced from
synchrotron self-Compton emission of the X-ray flare photons with the
electrons which produced them. Should VHE emission be detected from a
GRB coincident with X-ray flares, the time profile of the VHE emission
could be used to distinguish between these two origins of the X-ray
flares.

No evidence for delayed VHE gamma-ray emission was seen from any of
the GRB locations observed here and upper limits have been placed on
the VHE emission at various times after the prompt GRB
emission. Although there are no reports of the detection of X-ray
flares or delayed X-ray emission from any of these GRBs, it is likely
that such emission was present in at least some of them given the
frequency with which it has been detected in GRBs observed by Swift.
Indeed, the light curve of GRB\,030329 shows large variability
amplitude a few days after the burst and, as shown in
Figure~\ref{GRB030329-lightcurve}, Whipple observations were taken
during these episodes. Apart from this, a measured redshift is only
available for one of the other bursts observed here and it is possible
that the remaining five occurred at distances too large to be
detectable in the VHE regime.

\citet{Soderberg:04} reported on an unusual GRB (GRB\,031203) that was
much less energetic than average. Its similarity, in terms of
brightness, to an earlier GRB (GRB\,980425) suggests that the nearest
and most common GRB events have not been detected up until now because
GRB detectors were not sensitive enough \citep{Sazonov04}. Most GRBs
that have been studied up until now lie at cosmological
distances. They generate a highly collimated beam of gamma rays
ensuring that they are powerful enough to be detectable at large
distances. Both of the less powerful GRBs detected to date occurred at
considerably lower redshifts; GRB\,980425 at {\it{z}}\,=\,0.0085 and
GRB\,031203 at {\it{z}}\,=\,0.1055. Although \citet{Soderberg:04}
conclude that up until now, GRB detectors have only detected the
brightest GRBs and that the nearest and most common GRB events have
been missed because they are less highly collimated and energetic,
\citet{Ramierz:05} argue that the observations of GRB\,031203 can
indeed be the result of off-axis viewing of a typical, powerful GRB
with a jet. Should future observations prove there to be a closer,
less powerful population of GRBs, these would be prime targets for
IACTs.

In the past year, the Whipple Observatory 10m Telescope has been used
to carry out follow-up observations on a number of GRBs detected by
the Swift GRB Explorer. The analysis of these observations will be the
subject of a separate paper \citep{Dowdall}.

The Very Energetic Radiation Imaging Telescope Array System (VERITAS)
is currently under construction at the Fred Lawrence Whipple
Observatory in Southern Arizona. Two of the four telescopes are fully
operational and it is anticipated that the four-telescope array will
be operational by the end of 2006. GRB observations will receive high
priority and, when a GRB notification is received, their rapid
follow-up will take precedence over all other observations. The
VERITAS Telescopes can slew at 1$^\circ$\,s$^{-1}$ thus enabling them
to reach any part of the visible sky in less than 3 minutes. When an
acceptable (i.e. at high enough elevation) GRB notification is
received during observing at VERITAS, an alarm sounds to alert the
observer that a GRB position has arrived. Upon receiving authorization
from the observer, the telescope slews immediately to the position and
data-taking begins. Given that the maximum time to slew to a GRB is 3
minutes, and that Swift notifications can arrive within 30\,s of the
GRB, it is possible that VERITAS observations could begin as rapidly
as 2-4 minutes after the GRB, depending on its location with respect
to the previous VERITAS target.

\begin{figure}
\plotone{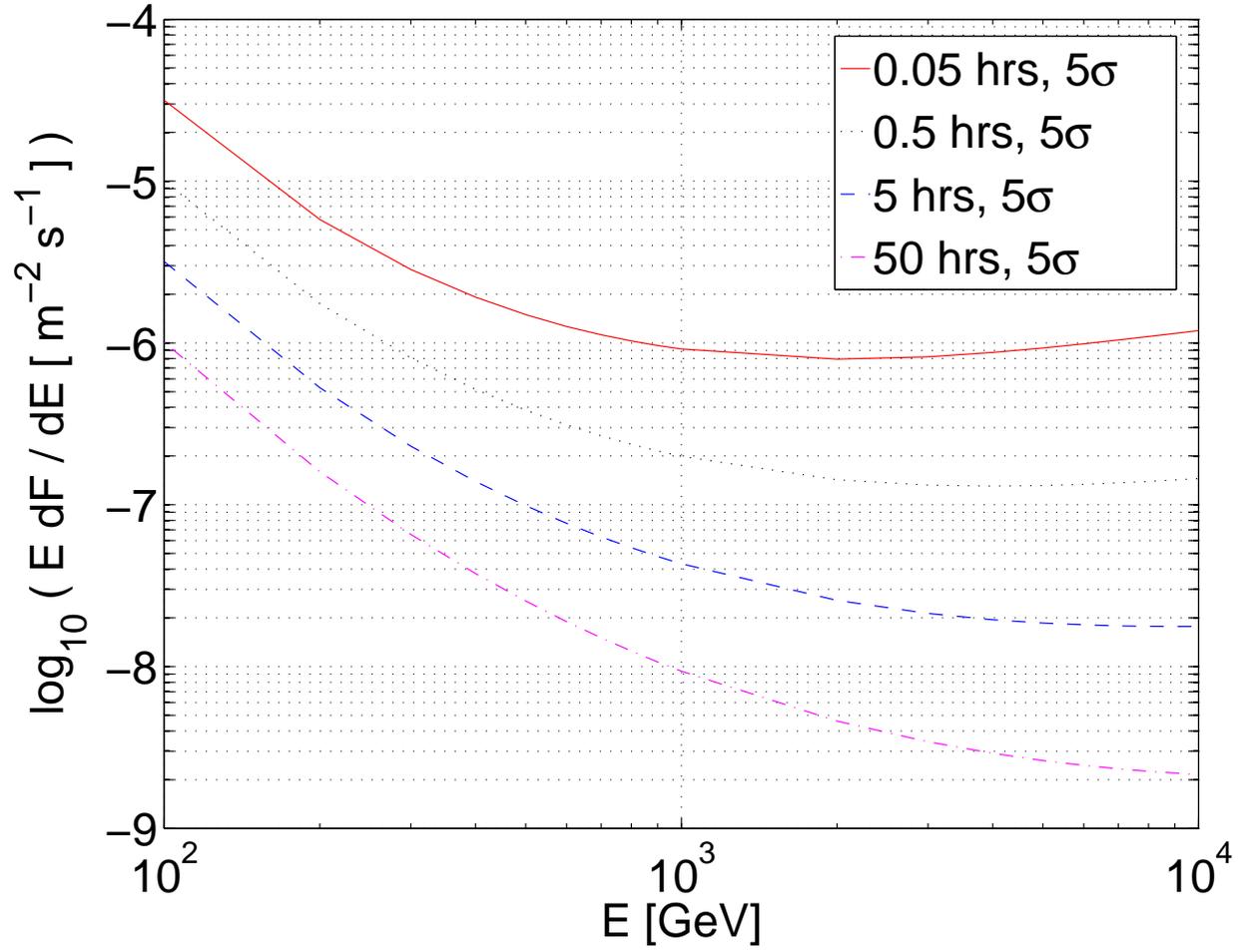}
\caption{\label{VERITAS-sensitivity}The sensitivity of the VERITAS
array for exposures of 50 hours, 5 hours, 0.5 hours, and 0.05 hours
(i.e., 3 minutes).}
\end{figure}

As has been shown above, the Whipple 10\,m Telescope is sensitive
enough to detect the GRB afterglow emission predicted by many
authors. With its improved background rejection and greater energy
range, VERITAS will be significantly more sensitive for GRB
observations than the Whipple 10m Telescope. The VERITAS sensitivity
for observations of different durations is shown in
Figure~\ref{VERITAS-sensitivity}. Based on the assumed rate of Swift
detections (100 year$^{-1}$), the fraction of sky available to
VERITAS, the duty-cycle at its site and the sun avoidance pointing of
Swift which maximizes its overlap with nighttime observations, it is
anticipated that $\sim$\,10 Swift GRBs will be observable each year with
VERITAS.

\section{Acknowledgements}
The authors would like to thank Emmet Roache, Joe Melnick, Kevin
Harris, Edward Little, and all of the staff at the Whipple Observatory
for their support. The authors also thank the anonymous referee for
his/her comments which were very useful and improved the paper. This
research was supported in part by the U. S. Department of Energy, the
National Science Foundation, PPARC, and Enterprise Ireland. Extensive
use was made of the GCN web pages (http://gcn.gsfc.nasa.gov/). The web
pages of Joachim Greiner and Stephen Holland
(http://www.mpe.mpg.de/$^\sim$jcg/grbgen.html and\\
http://lheawww.gsfc.nasa.gov/$^\sim$sholland/grb/index.html) proved
very useful in tracking down references and information related to the
GRBs discussed in this paper.


\clearpage

\begin{deluxetable}{ccccccc}
\tablewidth{0pt}
\tablecaption{The properties of the gamma-ray bursts described in this work. \label{grb_summary}}

\tablehead{  &\colhead{Discovery}&\colhead{Trigger}&                        &\colhead{Fluence\tablenotemark{a}}&\colhead{T90\tablenotemark{b}}      &\colhead{Energy band}\\
\colhead{GRB}&\colhead{Satellite}&\colhead{Number} &\colhead{z}             & \colhead{(erg\,cm$^{-2}$)}       &\colhead{(s)}                       &\colhead{(keV)}}
\startdata
021112       & HETE-2            & 2448            & ---                    & 2.1\,x\,10$^{-7}$                &  6.39                              & 30\,-\,400        \\ 
021204       & HETE-2            & 2486            & ---                    & ---                              &  ---                               & ---               \\ 
021211       & HETE-2            & 2493            & 1.006\tablenotemark{c} & 2.4\,x\,10$^{-6}$                &  2.80                              & 30\,-\,400        \\ 
030329       & HETE-2            & 2652            & 0.17\tablenotemark{d}  & 1.1\,x\,10$^{-4}$                &  22.76                             & 30\,-\,400        \\ 
030501       & INTEGRAL          & 596             & ---                    & 1.1\,x\,10$^{-6}$                &$\sim$\,75\tablenotemark{e}         & 25\,-\,100        \\ 
031026       & HETE-2            & 2882            & 6.67\tablenotemark{f}  & 2.8\,x\,10$^{-6}$                & 31.97                              & 30\,-\,400        \\ 
040422       & INTEGRAL          & 1758            & ---                    & ---\,\tablenotemark{g}           & 8\tablenotemark{h}                 & ---               \\ 
\enddata
\tablenotetext{a}{The fluence, where available, is quoted for the
energy range given in column\,7 over the duration listed in column
6. For most HETE-2 bursts, this was found at:
http://space.mit.edu/HETE/Bursts/Data.}
\tablenotetext{b}{Except for where a footnote is referenced, the
durations in this column are T90, the time interval during which 90\%
of the GRB photons were detected in the 30\,-\,400 keV energy band.}
\tablenotetext{c}{\citet{GCN1785:Redshift021211}.}
\tablenotetext{d}{\citet{GCN2020:VLT}.}
\tablenotetext{e}{The fluence and duration given in the table are from
burst observations with the Ulysses satellite and the SPI-ACS
instrument on INTEGRAL. The event was quite weak so there is a factor
of 2 uncertainty in the numbers quoted \citep{GCN2187:ipn}.
Observations with the IBIS/ISGRI instrument on INTEGRAL alone gave a
duration of $\sim$\,40 seconds for the burst
\citep{GCN2183:integral}.}
\tablenotetext{f}{This redshift was determined using
the redshift estimator described in \citet{Pelangeon:06}.}
\tablenotetext{g}{The fluence was not quoted for this burst over its
8 second duration. It had a fluence of
2.5\,x\,10$^{-7}$\,erg\,cm$^{-2}$ when integrated over
1 second\citep{GCN2572:integral}.}
\tablenotetext{h}{It was not stated by \citet{GCN2572:integral}
whether or not this duration is T90.}
\end{deluxetable}

\clearpage

\begin{deluxetable}{cccccc}
\tablewidth{0pt}
\tablecaption{The VHE GRB observations. \label{observations_and_results}}
\tablehead{  &\colhead{T$_{GRB}$\,-\,T$_{OBS}$\tablenotemark{a}} &\colhead{Exposure}&\colhead{Position Offset\tablenotemark{b}}&\colhead{T$_{GRB}$\,-\,T$_{UL}$\tablenotemark{c}}&\colhead{Flux\tablenotemark{d}}\\
\colhead{GRB}&\colhead{(hr)}                                    &\colhead{(min)}  &\colhead{(deg.)} &\colhead{(hr)} &\colhead{(Crab)}}
\startdata
021112       &  4.24                                             & 110.56           & 0.013 &   5.1        & $<$\,0.200                 \\
             & 28.63                                             &  55.28           & 0.013 &  29.0        & $<$\,0.303                 \\
021204       & 16.91                                             &  55.34           & 0.009 &  17.4        & $<$\,0.331                 \\
021211       & 20.69                                             &  82.79           & 0.058 &  21.9        & $<$\,0.325                 \\
030329       & 64.55                                             &  65.21           & 0.060 &  66.2        & $<$\,0.360                 \\
             & 112.58                                            &  83.17           & 0.022 & 113.8        & $<$\,0.279                 \\
             & 136.23                                            &  37.55           & 0.022 & 137.0        & $<$\,0.323                 \\
             & 162.14                                            &  27.74           & 0.022 & 162.4        & $<$\,0.519                 \\
             & 186.16                                            &  27.73           & 0.022 & 186.4        & $<$\,0.399                 \\
030501       &  6.58                                             &  83.10           & 0.001 &   7.3        & $<$\,0.265                 \\
031026       &  3.68                                             &  82.70           & 0.007 &   4.9        & $<$\,0.406                 \\
040422       &  3.99                                             &  27.63           & 0.062 &   4.2        & $<$\,0.620                 \\
\enddata

\tablenotetext{a}{The time in hours between the start of the GRB and
the beginning of observations with the Whipple 10m Telescope.}
\tablenotetext{b}{The angular separation between the position at which
these data were taken and the refined location of the GRB.}
\tablenotetext{c}{The length of time after the GRB for which the upper
limits (ULs) are quoted. Since all data are combined to compute the
upper limit, the mean time of the observations is quoted as the time
to which the upper limit pertains.}
\tablenotetext{d}{This is the flux upper limit in units of equivalent
Crab flux above the peak response energy of $\sim$ 400 GeV. Above this
energy, the integral Crab flux is
8.412\,$\pm$\,1.840\,x\,10$^{-11}$\,cm$^{-2}$\,s$^{-1}$.}
\end{deluxetable}

\clearpage

\clearpage

\clearpage

\clearpage

\end{document}